\documentclass[12pt,a4paper]{article}
\usepackage{amsfonts}
\usepackage{amssymb}
\usepackage{amsmath}
\usepackage{latexsym}
\begin{document}

\begin{center}
{\Large \bf Hierarchy problem and the cosmological constant in\\
\vspace{2mm} a five-dimensional Brans-Dicke brane world model} \\

\vspace{4mm}

Mikhail N.~Smolyakov\\
\vspace{0.5cm} Skobeltsyn Institute of Nuclear Physics, Moscow
State University,
\\ 119991, Moscow, Russia\\
\end{center}

\begin{abstract}
We discuss a new solution, admitting the existence of $dS_{4}$
branes, in five-dimensional Brans-Dicke theory. It is shown that,
due to a special form of a bulk scalar field potential, for
certain values of the model parameters the effective cosmological
constant can be made small on the brane, where the hierarchy
problem of gravitational interaction is solved. We also discuss
new stabilization mechanism which is based on the use of auxiliary
fields.
\end{abstract}

\section{Introduction}
Brane world models and their phenomenology have been widely
discussed in the last years. The most known models -- the
Arkani-Hamed, Dimopoulos and Dvali scenario \cite{ADD} and the
Randall-Sundrum model \cite{Sun}, -- provide elegant, although
different, solutions to the hierarchy problem of gravitational
interaction. It seems that the second model is more consistent,
because it takes into account the proper gravitational field of
the branes. Meanwhile it was shown that the four-dimensional
effective theory on the branes in the Randall-Sundrum model
contains a massless scalar field -- the radion, which is a
consequence of the fact that the distance between the branes is
not fixed by the parameters of the model. The coupling constant of
this field to matter on the negative tension brane, which is
assumed to trap the Standard Model fields, appears to be very
large, which contradicts experimental data even at the level of
classical experiments \cite{Rubakov,boos}.

This problem was solved by introducing an extra scalar field
living in the bulk. The most consistent model was proposed in
paper \cite{wolfe}, where exact solutions to equations of motion
for the background metric and the scalar field were found. The
size of the extra dimension is defined by the boundary conditions
for the scalar field on the branes.

The models discussed above assume the metric on the branes to be
the flat Minkowski metric. At the same time it is evident that
more realistic models should account for a cosmological evolution
on the branes. This problem is widely discussed in scientific
literature, see, for example, papers \cite{wolfe,Barger:2000wj},
reviews \cite{Brax:2003fv,Brax:2004xh} and references therein.
Quite an interesting class of the brane world models is the one
describing background solutions with $dS_{4}$ metric on the
branes. There are examples of such solutions in a slightly
modified Randall-Sundrum model (with non-equal brane tensions)
\cite{Kaloper:1999sm,Karch:2000ct}, as well as in models with
additional matter on the branes and in the bulk
\cite{Binetruy:1999hy} including scalar field living in the bulk
\cite{Cline:2000ky,Kanti:2002ta}. The latter models are of
particular interest, because an additional scalar field can fix
the size of the extra dimension thus giving stabilized models.

One of the standard scalar-tensor theories of gravity is the
Brans-Dicke theory (see \cite{Misner,Weinberg}). In the context of
brane world models this theory was also discussed in the
literature for the case of static \cite{mmsv} and time dependent
\cite{eqm,Mendes:2000wu} solutions, including background solutions
with $dS_{4}$ metric on the branes. Although one can transform the
theory to the Einstein frame, in which the scalar field minimally
couples to gravity, the theory in the original form can provide
elegant and simple background solutions. Moreover, in principle it
is possible that we live in the world, in which five-dimensional
scalar field is non-minimally coupled to five-dimensional
curvature (this is defined by the interaction of matter on the
branes with gravity, i.e. by the metric which we are supposed to
perceive). In this case it is more convenient to consider
untransformed action in the Jordan frame, in which the scalar
field non-minimally couples to gravity.

As it was mentioned above, brane world models are of particular
interest because they provide elegant solution to the hierarchy
problem of gravitational interaction. Nevertheless, such models
should also describe cosmological evolution at least on the late
stages. In this paper we discuss a stabilized brane world model in
five-dimensional Brans-Dicke theory admitting $dS_{4}$ branes. We
also discuss the values of fundamental parameters, which can make
the effective cosmological constant on the brane which is supposed
to contain SM fields (and where the hierarchy problem of
gravitational interaction is solved) very small.

\section{The model}
Let us consider gravity in a five-dimensional space-time
$E=M_4\times S^1/Z_2$, interacting with two branes and with the
scalar field $\phi$. Let us denote coordinates in $E$ by
$\{x^M\}=\{t,x^i,y\}$, $ M=0,1,2,3,4$, where $x^{0}\equiv t$;
$\{x^i\},\: i=1,2,3$ are three-dimensional spatial coordinates and
the coordinate $y\equiv x^4$, $-L\le y\le L$, corresponds to the
extra dimension. The extra dimension forms the orbifold $S^1/Z_2$,
which is a circle of diameter $2L/\pi$ with the points $y$ and
$-y$ identified. Correspondingly, the metric $g_{MN}$ and the
scalar field $\phi$ satisfy the orbifold symmetry conditions
\begin{eqnarray}\label{eq} g_{\mu\nu}(x,-y)=g_{\mu\nu}(x,y),
\quad g_{\mu4}(x,-y)=-g_{\mu4}(x,y), \\ \nonumber
g_{44}(x,-y)=g_{44}(x,y), \quad \phi(x,-y) =\phi(x,y),
\end{eqnarray}
$\mu=0,1,2,3$. The branes are located at the fixed points of the
orbifold $y=0$ è $y=L$.

The action of the model has the form
\begin{eqnarray}\label{s}
S=\int d^4x \int_{-L}^{L}dy \sqrt{-g}\left[\phi
R-\frac{\omega}{\phi}g^{MN}\partial_M\phi
\partial_N\phi-V(\phi)\right]-\\ \nonumber
-\int_{y=0}\sqrt{-\tilde
g}\lambda_1(\phi)d^4x-\int_{y=L}\sqrt{-\tilde
g}\lambda_2(\phi)d^4x.
\end{eqnarray}

Here $V(\phi)$ is the scalar field potential in five-dimensional
space-time, $\lambda_{1,2}(\phi)$ are scalar field potentials on
the branes, $\omega$ is the five-dimensional Brans-Dicke parameter
(we suppose that $\omega\gg 1$), $\tilde g_{\mu\nu}$ denotes
induced metric on the branes. The signature of the metric $g_{MN}$
is chosen to be $(-,+,+,+,+)$. Subscripts 1 and 2 label the
branes. We also note that the dimension of the field $\phi$ is
$[mass]^{3}$.

We consider the following standard form of the background metric,
which is often used in brane world models (see, for example,
\cite{wolfe})
\begin{equation}\label{backgmetric}
ds^2=\gamma_{MN}dx^{M}dx^{N}=e^{-2A(y)}\left(-dt^{2}+a^{2}(t)\eta_{ij}dx^{i}dx^{j}\right)+dy^2
\end{equation}
with $\eta_{ij}=diag(1,1,1)$, and the following form of the
background solution for the scalar field
\begin{equation}\label{backgscalar}
\phi(x,y)=\phi(y)
\end{equation}
(the background solution is the solution corresponding to the
vacuum states on the branes, i.e. when all the fields on the
branes are in their vacuum states).

In paper \cite{mmsv} a model, admitting a simple background
solution in the case $a(t)\equiv 1$, was proposed. Indeed, with
\begin{equation}\label{bulk-V-RS}
V(\phi)=\Lambda\phi
\end{equation}
the background solution for the metric and scalar field takes a
simple form \cite{mmsv}:
$$A(y)=k|y|-kL,$$
\begin{equation}\label{phi-RS}
\phi=v\left(e^{-A}\right)^{\frac{1}{\omega+1}},
\end{equation}
where $L$ is the size of the extra dimension, which is defined by
boundary conditions on the branes together with constant $v$, and
$k$ is given by
\begin{equation}\label{k-flat}
k^{2}=-\Lambda\frac{\left(\omega+1\right)^{2}}{\left(3\omega+4\right)\left(4\omega+5\right)}.
\end{equation}
Parameter $\Lambda$ (and consequently parameter $k$) characterizes
the energy scale of five-dimensional gravity.

It appears that it is possible to modify slightly the bulk
potential (\ref{bulk-V-RS}) to get $dS_{4}$ background metric on
the branes. We consider ansatz (\ref{backgmetric}),
(\ref{backgscalar}) for the vacuum background solution with
\begin{equation}\label{backa}
a(t)=e^{Ht},
\end{equation}
\begin{equation}\label{backphi}
\phi=v\left(e^{-A}\right)^{\frac{1}{\omega+1}},
\end{equation}
where $H$ is the four-dimensional Hubble parameter on the branes,
$v$ is a constant. The Hubble parameter $H$ and the constant $v$
will acquire their values after solving the corresponding
equations of motion. Note that the form of the background solution
for the scalar field is the same as the one used in \cite{mmsv}
for the case $a(t)\equiv 1$ (see equation (\ref{phi-RS})).
Assumption (\ref{backphi}) simplifies considerably equations of
motion for the scalar field and the Einstein equations following
from (\ref{s}).

The Einstein equations for a general form of the metric and the
equation of motion for the Brans-Dicke scalar field can be found,
for example, in \cite{eqm} (these equations are obtained for
$\phi=\varphi^{2}/\left(8\pi\right)$ and $\lambda_{1,2}(\phi)=0$,
nevertheless, the case $\lambda_{1,2}(\phi)\ne 0$ can be easily
restored). These equations have quite a complicated form. But for
our choice of the background solution (\ref{backgmetric}),
(\ref{backa}) and (\ref{backphi}) the corresponding equations take
a simpler form:
\begin{enumerate}
\item $\mu\nu$-component
\begin{eqnarray}\label{eqmunu}
6H^2e^{2A}+\frac{6\omega+8}{\omega+1}A''-{A'}^{2}\frac{12\omega^{2}+31\omega+20}{\left(\omega+1\right)^{2}}=\\
\nonumber =\frac{V(\phi)}{\phi}+
\frac{\lambda_{1}(\phi)}{\phi}\delta(y)+\frac{\lambda_{2}(\phi)}{\phi}\delta(y-L),
\end{eqnarray}

\item $44$-component
\begin{eqnarray}\label{eq44}
12H^2e^{2A}-{A'}^{2}\frac{12\omega^{2}+31\omega+20}{\left(\omega+1\right)^{2}}=\frac{V(\phi)}{\phi},
\end{eqnarray}

\item equation for the field $\phi$
\begin{eqnarray}\label{eqf}
12H^2e^{2A}+\frac{6\omega+8}{\omega+1}A''-{A'}^{2}\frac{12\omega^{2}+31\omega+20}{\left(\omega+1\right)^{2}}=\\
\nonumber =\frac{dV(\phi)}{d\phi}+
\frac{d\lambda_{1}(\phi)}{d\phi}\delta(y)+\frac{d\lambda_{2}(\phi)}{d\phi}\delta(y-L),
\end{eqnarray}
where $'=\frac{d}{dy}$.
\end{enumerate}

We suppose that the bulk scalar field potential has the form
\begin{equation}\label{bulk-phi-V}
V(\phi)=\Lambda\phi+\frac{\beta}{\phi^{2\omega+1}},
\end{equation}
where $\Lambda<0$, $\beta<0$. The physical motivation of such a
choice of the potential is the following: the first term is
responsible for the solution to the hierarchy problem of
gravitational interaction on the brane at $y=L$, whereas the
second term is responsible for the constant Hubble parameter on
the branes (i.e. $dS_{4}$ space-time on the branes). One can see
that the second term in (\ref{bulk-phi-V}) is fine-tuned because
the five-dimensional Brans-Dicke parameter $\omega$ is utilized in
the definition of potential (\ref{bulk-phi-V}). This fine-tuning
of the five-dimensional bulk scalar field potential is similar to
the choice of cosmological constant in four-dimensional gravity to
get $dS_{4}$ space-time. Note that the vacuum structure of the
four-dimensional space-time on the branes is defined by the bulk
scalar field potential, not by the potentials on the branes. We
also suppose that parameters $\beta$, $\Lambda$ are in the $TeV$
energy range and do not contain hierarchical difference.

First, let us consider equations (\ref{eqmunu}), (\ref{eq44}) and
(\ref{eqf}) in the bulk. The corresponding solution to these
equations of motion has the form
\begin{equation}
A=-\ln{\left(C_{1}e^{k|y|}+C_{2}e^{-k|y|}\right)}
\end{equation}
with $k$ defined by equation (\ref{k-flat}). We also can get
\begin{equation}\label{cc0}
H^{2}=-4k^{2}\frac{3\omega+4}{3\omega+3}C_{1}C_{2}
\end{equation}
and
\begin{equation}\label{Hubble}
H^{2}=-\beta\frac{\left(\omega+1\right)}{3v^{2\omega+2}},
\end{equation}
which means that the four-dimensional Hubble parameter is
expressed through the parameter $\beta$ of the bulk scalar field
potential, the five-dimensional Brans-Dicke parameter $\omega$ and
the constant $v$, which will be defined below.

Let us discuss the values of the constants $C_{1}$, $C_{2}$. We
are interested in the effective theory on the brane at $y=L$ (it
will be shown below that the hierarchy problem is solved on this
brane). To this end we should take such $C_{1}$, $C_{2}$ that make
the four-dimensional coordinates on this brane Galilean (see
\cite{Rubakov,boos,Kubyshin}), i.e. $A|_{y=L}=0$. Thus,
\begin{equation}\label{cc}
C_{1}e^{kL}+C_{2}e^{-kL}=1.
\end{equation}
Equations (\ref{cc0}) and (\ref{cc}) define the values of $C_{1}$,
$C_{2}$ (we suppose, that the Hubble parameter $H\ll k$ because
$k$ is in the $TeV$ range, whereas $H$ should correspond to the
late time accelerated expansion).
\begin{equation}
C_{2}=e^{kL}\frac{1+\sqrt{1+\frac{(3\omega+3)H^{2}}{(3\omega+4)k^{2}}}}{2}\approx
e^{kL}\left(1+\frac{3(\omega+1)H^{2}}{4(3\omega+4)k^{2}}\right),
\end{equation}
\begin{equation}
C_{1}\approx -e^{-kL}\frac{3(\omega+1)H^{2}}{4(3\omega+4)k^{2}},
\end{equation}
which results in
\begin{eqnarray}\label{backgr-metric}
e^{-A}&\approx &e^{(kL-k|y|)}+\frac{3(\omega+1)H^{2}}{2(3\omega+4)k^{2}}\sinh{\left(kL-k|y|\right)}=\\
\nonumber &=&e^{(kL-k|y|)}+\frac{(4\omega+5)\beta}{2\Lambda
v^{2\omega+2}}\sinh{\left(kL-k|y|\right)}.
\end{eqnarray}

Now let us turn to the boundary conditions on the branes. They can
be easily obtained from (\ref{eqmunu}), (\ref{eqf}) by the
standard procedure (see, for example, \cite{wolfe}) and have the
form
\begin{eqnarray}\label{bc1}
2\frac{6\omega+8}{\omega+1}A'|_{y=+0}&=&\frac{\lambda_{1}(\phi)}{\phi}\left|_{\phi=\phi(0)}\right.,\\
\nonumber
2\frac{6\omega+8}{\omega+1}A'|_{y=L-0}&=&-\frac{\lambda_{2}(\phi)}{\phi}|_{\phi=\phi(L)},
\end{eqnarray}
\begin{eqnarray}\label{bc2}
2\frac{6\omega+8}{\omega+1}A'|_{y=+0}&=&\frac{d\lambda_{1}(\phi)}{d\phi}|_{\phi=\phi(0)},\\
\nonumber
2\frac{6\omega+8}{\omega+1}A'|_{y=L-0}&=&-\frac{d\lambda_{2}(\phi)}{d\phi}|_{\phi=\phi(L)}.
\end{eqnarray}
Note that there are no boundary conditions coming from equation
(\ref{eq44}) because it does not contain terms $\sim A''$ and
$\sim \delta(y)$.

We will consider the following form of the potential on the branes
\begin{eqnarray}\label{brane-pot}
\lambda_{1,2}(\phi)=\pm
4\sqrt{3\omega+4}\sqrt{-\frac{\Lambda\phi^2}{4\omega+5}-\frac{\beta}{\phi^{2\omega}}}+F_{1,2}(x)\left(\phi-\phi_{1,2}\right),
\end{eqnarray}
where $F_{1,2}(x)$ are scalar fields and $\phi_{1,2}$ are
constants. The absence of the kinetic terms for the fields
$F_{1,2}(x)$ looks rather strange. Nevertheless one can recall
that supersymmetry is based on the use of such "auxiliary"\,
fields, which are necessary for reaching the closure of the
supersymmetry algebra \cite{Ramond}. A simple example with the
fields of such type in classical field theory can be also found in
\cite{Ramond}.

Of course, one can use a more conventional form of the stabilizing
potentials, for example
\begin{eqnarray}\label{brane-pot-conv}
\lambda_{1}(\phi)&=&4\sqrt{3\omega+4}\sqrt{-\frac{\Lambda\phi^2}{4\omega+5}
-\frac{\beta}{\phi^{2\omega}}}-\frac{4(\omega+1)\sqrt{3\omega+4}\beta}
{\sqrt{-\frac{\Lambda\phi_{1}^{4\omega+4}}{4\omega+5}-\beta\phi_{1}^{2\omega+2}}}
\left(\phi-\phi_{1}\right)+\\ \nonumber
&+&q_{1}^{2}\left(\phi-\phi_{1}\right)^{2}, \\
\label{brane-pot-conv1}
\lambda_{2}(\phi)&=&-4\sqrt{3\omega+4}\sqrt{-\frac{\Lambda\phi^2}{4\omega+5}
-\frac{\beta}{\phi^{2\omega}}}+\frac{4(\omega+1)\sqrt{3\omega+4}\beta}{\sqrt{-\frac{\Lambda\phi_{2}^{4\omega+4}}
{4\omega+5}-\beta\phi_{2}^{2\omega+2}}}\left(\phi-\phi_{2}\right)+\\
\nonumber &+&q_{2}^{2}\left(\phi-\phi_{2}\right)^{2}.
\end{eqnarray}
The terms $\sim q_{1,2}^2$ usually are introduced to ensure the
absence of tachyonic modes in the linearized theory \cite{mmsv1}.
But such form of the potentials appears to be highly fine-tuned.
One can see that there is strong relation between the terms even
if we consider only one of the brane potentials
((\ref{brane-pot-conv}) or (\ref{brane-pot-conv1})). The auxiliary
fields add extra degrees of freedom, which make the inner
fine-tuning of the terms in the brane potentials unnecessary.
Nevertheless, there remains fine-tuning in (\ref{brane-pot}) --
the first term in (\ref{brane-pot}) is defined by the form of the
bulk scalar field potential and by the form of the background
metric (\ref{backgmetric}) (one can check it by straightforward
calculations). Such fine-tuning is inherent to almost all brane
world models with compact extra dimension and two branes (see, for
example, \cite{Sun,wolfe}).

Equations of motions for fields $F_{1,2}(x)$ give (which can be
obtained by means of the standard variation procedure with respect
to the fields $F_{1,2}(x)$)
\begin{eqnarray}\label{brane-phi1}
\phi|_{y=0}=\phi_{1},\\ \label{brane-phi2} \phi|_{y=L}=\phi_{2}.
\end{eqnarray}
We see that these equations do not contain fields $F_{1,2}(x)$
itself. Equations (\ref{bc1}) appear to be satisfied automatically
for the choice (\ref{brane-pot}) (one can check it using
(\ref{eq44}), (\ref{bulk-phi-V}) and (\ref{Hubble})), whereas
equations (\ref{bc2}) define the background values of the fields
$F_{1,2}(x)$
\begin{eqnarray}\label{F11}
F_{1}&=&-\frac{4(\omega+1)\sqrt{3\omega+4}\beta}{\sqrt{-\frac{\Lambda\phi_{1}^{4\omega+4}}{4\omega+5}-\beta\phi_{1}^{2\omega+2}}},\\
\label{F12}
F_{2}&=&\frac{4(\omega+1)\sqrt{3\omega+4}\beta}{\sqrt{-\frac{\Lambda\phi_{2}^{4\omega+4}}{4\omega+5}-\beta\phi_{2}^{2\omega+2}}}
\end{eqnarray}
(we see, that the fields $F_{1,2}(x)$ appear in equation of motion
for the scalar field $\phi$). Thus, although fields $F_{1,2}(x)$
do not contain kinetic terms and are not dynamical, they can be
treated as normal fields and one can use the standard variation
technique to obtain corresponding equations of motion
\cite{Ramond}.

Equations (\ref{brane-phi1}) and (\ref{brane-phi2}) define the
size of the extra dimension. Indeed, from (\ref{backphi}),
(\ref{backgr-metric}) and (\ref{brane-phi2}) it follows that
$v=\phi_{2}$. Then, neglecting the contribution of the term
proportional to $H^{2}/k^{2}$ (because we suppose that $H\ll k$),
we get from (\ref{backphi}) and (\ref{brane-phi1}) (see
\cite{mmsv} for details)
\begin{equation}\label{edsize}
L\approx\frac{\left(\omega+1\right)}{k}\ln{\left(\frac{\phi_{1}}{\phi_{2}}\right)}.
\end{equation}
Note that the same relation was obtained for the simpler model
discussed in \cite{mmsv}. The four-dimensional Planck mass on the
brane at $y=L$ can also be easily obtained. For our purposes the
contribution proportional to $H^{2}/k^{2}$ in (\ref{backphi}),
(\ref{backgr-metric}) can be also neglected, and in this
approximation the corresponding Planck mass is (detailed
derivation can be found in \cite{mmsv})
\begin{equation}\label{Planck1}
M^{2}_{Pl}=\frac{1}{2}\int_{-L}^{L}\phi e^{-2A}dy\approx
\frac{\phi_{1}}{2k}e^{2kL}.
\end{equation}
Thus, if $kL\approx 35$ and parameters $\phi_{1}$, $k$ of the
theory lie in the $TeV$ range, the hierarchy problem on the brane
at $y=L$ is solved in the way analogous to that used in the
original Randall-Sundrum model \cite{Sun}. Thus, we get weak
four-dimensional gravity on the brane at $y=L$ with $M_{Pl}\sim
10^{19}\,GeV$, whereas five-dimensional gravity is characterized
by the $TeV$ energy scale.

The five-dimensional Hubble parameter on the brane is defined by
equation (\ref{Hubble}), which can be rewritten as
\begin{equation}\label{Hubble1}
H^{2}=-\beta\frac{\left(\omega+1\right)}{3\phi_{2}^{2\omega+2}}.
\end{equation}
If we suppose that $\beta\approx -1\,TeV^{6\omega+8}$,
$\omega\approx 45$ and $\phi_{2}\approx 10\, TeV^{3}$, then the
Hubble parameter in (\ref{Hubble1}) has the same small value as
that in ordinary four-dimensional gravity defined by vacuum energy
density
\begin{equation}\label{Hubble2}
\rho_{\Lambda}\sim H^{2}M_{Pl}^{2}\sim 10^{-47}GeV^{4}.
\end{equation}
Such value of $H$ can correspond to the late time accelerated
expansion of the Universe if we suppose that our four-dimensional
world lives on the brane at $y=L$.

Note, that such a small value of the effective cosmological
constant appears because of the large power $\sim 2\omega$ in the
denominator of the fine-tuned bulk potential (\ref{bulk-phi-V}).
One can argue that such a way of obtaining a small effective
cosmological constant is analogous to introducing its small value
"by hand". It is not the case in the model discussed above. It is
somewhat similar to the solution of the hierarchy problem of
gravitational interaction in the Arkani-Hamed, Dimopoulos and
Dvali scenario \cite{ADD}, where four-dimensional Planck mass on
the brane appears to be large due to a large number of extra
dimensions. But it is necessary to note that the large value
$10^{2\omega+2}$ is not introduced to the model by hand. Indeed,
if $\phi_{2}\approx 1\, TeV^{3}$, then from (\ref{Hubble1}) it
follows that $H^{2}\approx 15\,TeV^{2}$, which is extremely large
value in comparison with the present day value of the Hubble
parameter. Thus, the effective four-dimensional Hubble parameter
on the brane depends on the value of the scalar field $\phi$ on
the brane. The value $\phi_{2}=10\,TeV^{3}$ does not create a new
hierarchy itself.

\section{Stability}
A consistent study of stability of our model, at least in the
linear approximation, appears to be quite a complicated task,
which goes beyond the scope of this paper. Indeed, one should
derive linearized equations of motion above the background
solution, isolate the physical degrees of freedom and find the
mass spectra of the excitations. Nevertheless we can simplify the
problem and consider the simpler case $\beta=0$, for which we can
show that our method of fixing the size of the extra dimension by
utilizing the auxiliary fields $F_{1,2}(x)$ does not lead to
instabilities at least in the simplest cases. For $\beta=0$ the
potentials on the branes and the bulk potential take the form
(compare with those used in \cite{mmsv}):
\begin{eqnarray}\label{brane-pot1}
V(\phi)=\Lambda \phi,\qquad \lambda_{1,2}(\phi)=\pm
4\sqrt{-\Lambda}\sqrt{\frac{3\omega+4}{4\omega+5}}\phi+F_{1,2}(x)\left(\phi-\phi_{1,2}\right).
\end{eqnarray}
Boundary conditions (\ref{bc1}) and (\ref{bc2}) lead to the
following background values
\begin{eqnarray}
F_{1}&\equiv&0,\\
F_{2}&\equiv&0,
\end{eqnarray}
which also follow from (\ref{F11}), (\ref{F12}) for $\beta=0$.

Linearized gravity in five-dimensional Brans-Dicke stabilized
brane world models was thoroughly examined in \cite{mmsv1}. In
particular, the model with
\begin{eqnarray}\label{super}
& V(\phi)=\Lambda \phi,\qquad
\lambda_{1,2}=\pm4\sqrt{-\Lambda}\sqrt{\frac{3\omega+4}{4\omega+5}}\phi+\frac{\beta_{1,2}^{2}}{2}(\phi-v_{1,2})^2
\end{eqnarray}
was also considered. It was shown that this model is stable and
does not contain the scalar zero mode. The only difference between
(\ref{brane-pot1}) and (\ref{super}) is the stabilizing potentials
on the branes (such potentials are often called stabilizing
potentials because they define the values of the scalar field on
the branes, which leads to fixation of the size of the extra
dimension and thus to stabilized model). The results obtained in
\cite{mmsv1} show that from the four-dimensional point of view
linearized gravity can be described by the tensor (massless and
massive tensor gravitons) and scalar (massive scalar modes)
physical degrees of freedom. The tensor sector does not depend on
the form of the stabilizing potentials and, as it was shown in
\cite{mmsv1}, does not contain tachyons. But the scalar sector of
the model changes under the change of the stabilizing potentials.
We will use results of \cite{mmsv1} to show that the new method of
stabilization does not lead to any new unwanted consequences in
the scalar sector. We will not present here the full set of
linearized equations of motion because these equations are quite
tedious. One can find detailed calculations in \cite{mmsv1}.

To start with, let us parameterize the metric and the scalar field
as
\begin{eqnarray}\label{razlozh}
g_{MN}(x,y)&=&\gamma_{MN}(y)+h_{MN}(x,y),\\
\phi(x,y)&=&\phi_{0}(y)+f(x,y),\\
F_{1,2}(x)&=&F_{1,2}^{0}+j_{1,2}(x)=j_{1,2}(x).
\end{eqnarray}
For consistency with the previous sections below we will write
$F_{1,2}$ and $\phi(y)$ for the background solution instead of
$F_{1,2}^{0}$ and $\phi_{0}(y)$.

It was shown in \cite{mmsv1} that the physical degrees of freedom
of the scalar sector for any form of the bulk scalar field
potential $V(\phi)$ can be completely described by the new field
$g$:
$$g(x,y) =
e^{-2A}\phi^{2/3}\left(h_{44}(x,y)+\frac{2}{3}\frac{f(x,y)}{\phi}\right).$$
It was also shown that the fluctuations $f$ of the stabilizing
scalar field $\phi$ can be expressed in terms of the field $g$
through the gauge condition \cite{mmsv1}
\begin{equation}\label{scal-gauge}
g'=\frac{4}{3}\left(\omega+\frac{4}{3}\right)e^{-2A}\frac{\phi'}{\phi^{4/3}}f.
\end{equation}
The corresponding equation of motion for the field $g$ in the bulk
looks like \cite{mmsv1}
\begin{equation}\label{ug-S}
\left(g'\frac{\phi^{5/3}e^{2A}}{\phi'^{2}}\right)'
-\frac{2}{9}\left(3\omega+4\right)\frac{e^{2A}}{\phi^{1/3}}
g+\frac{\phi^{5/3}e^{2A}}{\phi'^{2}}\partial_\mu
\partial^\mu g =0.
\end{equation}

The difference between the model examined in \cite{mmsv1} and the
model under consideration is the brane scalar field potentials.
Indeed, from equations (\ref{brane-phi1}), (\ref{brane-phi2}) it
follows that $f|_{y=0}=f|_{y=L}=0$ and thus using
(\ref{scal-gauge}) we get boundary conditions
\begin{eqnarray}\label{bcc}
&g'|_{y=+0}=0, &\\ \nonumber &g'|_{y=L-0}=0.&
\end{eqnarray}
The latter conditions differ from those obtained in \cite{mmsv1}
for (\ref{super}).

There are also "boundary conditions", following from the
linearized equation of motion for the field $f$ \cite{mmsv1},
which, for our choice of the brane scalar field potentials
(\ref{brane-pot1}), take the form
\begin{eqnarray}\label{bccc}
&j_{1}(x)=-\frac{3\phi^{1/3}}{\phi'}e^{2A}\partial_\mu
\partial^\mu g|_{y=+0},
&\\ \nonumber
&j_{2}(x)=\frac{3\phi^{1/3}}{\phi'}e^{2A}\partial_\mu
\partial^\mu g|_{y=L-0}.&
\end{eqnarray}
We see from (\ref{bccc}) that the fields $j_{1}(x)$, $j_{2}(x)$
are not dynamical and appear to be defined by the values of the
field $g$ on the branes. The system appears to be not
overconstrained.

To make the mode decomposition, we represent $g(x,y)$ in a
standard way \cite{mmsv1}:
$$g(x,y)=\sum_{n=1}^{\infty}\varphi_{n}(x)g_{n}(y),\qquad
\eta^{\mu\nu}\partial_{\mu}\partial_{\nu}\varphi_{n}(x)=\mu_{n}^{2}\varphi_{n}(x).$$
Now the spectrum of scalar modes is defined by equations
\begin{equation}\label{ug-S1}
\left(g_{n}'\frac{\phi^{5/3}e^{2A}}{\phi'^{2}}\right)'
-\frac{2}{9}\left(3\omega+4\right)\frac{e^{2A}}{\phi^{1/3}}
g_{n}+\frac{\phi^{5/3}e^{4A}}{\phi'^{2}}\mu_n^2 g_{n} =0,
\end{equation}
\begin{eqnarray}\label{bcc1}
&g_{n}'|_{y=+0}=0, &\\ \nonumber &g_{n}'|_{y=L-0}=0,&
\end{eqnarray}
where $g_{n}(y)$ is the wave function of the four-dimensional mode
with the mass $\mu_{n}$ \cite{mmsv1}. It is not difficult to show
(multiplying (\ref{ug-S}) by $g_{n}$ and integrating it in the
limits $0<y<L$) that $\mu_{n}^2>0$. As for the zero mode with
$\mu_{0}=0$, its wave function $g_{0}\equiv 0$. Thus, there is no
zero scalar mode, which is inherent to stabilized brane world
models.

We showed that the model with $\beta=0$ is stable at least under
the small fluctuations of the fields and our method of fixing the
size of the extra dimension does not lead to instabilities. Other
properties of the spectrum, such as orthogonality of the wave
functions $g_{n}$, can be easily obtained using the results of
\cite{mmsv1}. As for the case $\beta\ne 0$, for a small $H$
corresponding to (\ref{Hubble2}) one also expects stability of the
model. Indeed, if all parameters of the model with $\beta=0$ lie
in the $TeV$ range, the masses of the lowest modes in the
four-dimensional effective theory on the brane are also expected
to lie in the $TeV$ range, as well as the mass gaps between the
modes. The case $\beta\ne 0$, providing the present day value of
$H$, in principle leads to the modification of the spectra of
tensor and scalar modes, but this modification can be neglected
because of the extremely small value of $H$ in comparison with
$TeV$ energy scale. In other words, we expect practically the same
tower of the modes (including the massless graviton) propagating
in the $dS_{4}$ space-time instead of the flat Minkowsky
space-time, and it does not pose any problems with stability (we
could expect ghost modes if the masses of the tensor modes were
smaller than $2H^2$, as it happens in four-dimensional massive
gravity \cite{Higuchi}, but it is not the case under
consideration). Locally we can even neglect the influence of the
non-zero Hubble parameter (for example, when considering the
collider phenomenology or Newtonian gravity). For these reasons
the model proposed in this paper seems to be stable.

\section{Conclusion}

In this paper we discussed stabilized brane world model in
five-dimensional Brans-Dicke theory. The choice of the bulk
potential was motivated by the demand to have $dS_{4}$ space-time
on the branes in the vacuum state (i.e. when there are no any
matter fields on the branes). We note, that our solution is
stationary, because the background solution for the scalar field
does not depend on time, the size of the extra dimension is fixed
and the four-dimensional Hubble parameter is constant. Such
situation is realized in the limit $x^{0}\to\infty$. Indeed, in
this case the ordinary and dark matter average densities on the
branes tend to zero, solutions for the metric and the scalar field
tend to this background solution.

Nevertheless the appropriate form of the bulk and brane scalar
field potential should be fine-tuned in order to get $dS_{4}$
space-time on the branes. As it was mentioned above, such
fine-tuning is the price we have to pay in order to get solutions
with desired properties. The fine-tuning of the bulk potential is
in some sense analogous to the choice of cosmological constant in
ordinary four-dimensional gravity in order to get maximally
symmetric $dS_{4}$ space-time, whereas the fine-tuning of the
brane potentials is necessary for the self-consistency of the
solution.

The advantages of the model presented above are the following.
\begin{enumerate}
\item
The hierarchy of gravitational interaction is solved in the model
in the way analogous to the one utilized in the original
Randall-Sundrum model \cite{Sun} (see equation (\ref{Planck1})).
\item
The size of the extra dimension is fixed (see equation
(\ref{edsize})).
\item
The small four-dimensional Hubble parameter is defined by the
parameter $\phi_{2}$ and can account for the late time accelerated
expansion on the brane.
\end{enumerate}

We note that the effective cosmological constant on the branes
appears to be small because of the large power in the denominator
of the second term in the bulk scalar field potential
(\ref{bulk-phi-V}), which is similar in some sense to the solution
of the hierarchy problem of gravitational interaction in the
Arkani-Hamed, Dimopoulos and Dvali scenario \cite{ADD}. As for the
fine-tuning of the scalar field potentials, in a most stabilized
five-dimensional brane world models one should use special forms
of the potential to get stationary vacuum solution, and the
five-dimensional Brans-Dicke theory is not an exception.

We hope that the results presented in this paper can be
interesting for a future investigations of brane world models.

\section*{Acknowledgments}

The author is grateful to I.P.~Volobuev for valuable discussions.
The work was supported by grant of Russian Ministry of Education
and Science NS-4142.2010.2, RFBR grant 08-02-92499-CNRSL-a and
state contract 02.740.11.0244.

\end{document}